\documentclass[conference]{IEEEtran}
\IEEEoverridecommandlockouts

\usepackage[dvipdfmx]{graphicx}
\usepackage{url}

\usepackage{multirow}

\usepackage{cite}
\usepackage{amsmath,amssymb,amsfonts}
\usepackage{algorithmic}
\usepackage{graphicx}
\usepackage{xcolor}
\def\BibTeX{{\rm B\kern-.05em{\sc i\kern-.025em b}\kern-.08em
 T\kern-.1667em\lower.7ex\hbox{E}\kern-.125emX}}
\begin{document}

\title{Instability of financial markets \\ by optimizing investment strategies \\ investigated by an agent-based model
\thanks{Note that the opinions contained herein are solely those of the authors and do not necessarily reflect those of SPARX Asset Management Co., Ltd.}
}

\author{
\IEEEauthorblockN{1\textsuperscript{st} Takanobu Mizuta}
\IEEEauthorblockA{\textit{SPARX Asset Management Co. Ltd.} \\ Tokyo, Japan \\ https://orcid.org/0000-0003-4329-0645}
\and
\IEEEauthorblockN{2\textsuperscript{nd} Isao Yagi}
\IEEEauthorblockA{\textit{Faculty of Informatics} \\
\textit{Kogakuin University} \\
Tokyo, Japan}
\and
\IEEEauthorblockN{3\textsuperscript{rd} Kosei Takashima}
\IEEEauthorblockA{\textit{Faculty of Economics and Business Administration} \\
\textit{Nagaoka University} \\
Niigata, Japan}
}

\maketitle

\IEEEpubidadjcol

\begin{abstract}
Most finance studies are discussed on the basis of several hypotheses, for example, investors rationally optimize their investment strategies. However, the hypotheses themselves are sometimes criticized. Market impacts, where trades of investors can impact and change market prices, making optimization impossible. In this study, we built an artificial market model by adding technical analysis strategy agents searching one optimized parameter to a whole simulation run to the prior model and investigated whether investors' inability to accurately estimate market impacts in their optimizations leads to optimization instability. In our results, the parameter of investment strategy never converged to a specific value but continued to change. This means that even if all other traders are fixed, only one investor will use backtesting to optimize his/her strategy, which leads to the time evolution of market prices becoming unstable. Optimization instability is one level higher than ``non-equilibrium of market prices.'' Therefore, the time evolution of market prices produced by investment strategies having such unstable parameters is highly unlikely to be predicted and have stable laws written by equations. This nature makes us suspect that financial markets include the principle of natural uniformity and indicates the difficulty of building an equation model explaining the time evolution of prices.
\end{abstract}

\begin{IEEEkeywords}
Optimization of investment strategy, Instability of financial market, Market impacts, Agent-based model, Multi-agent simulation, Artificial market model
\end{IEEEkeywords}

\section{Introduction}
Most finance studies are discussed on the basis of several hypotheses, for example, investors rationally optimize their investment strategies, markets are perfectly efficient, and so on. However, the hypotheses themselves are sometimes criticized. The rational optimization of investors' investment strategies is criticized in that investors themselves are not rational in the first place\cite{shiller2003efficient}. Also, optimization is impossible due to the limitations of the investors' observation range and calculation ability\cite{Shiozawa2019}. Market impacts, where trades of investors can impact and change market prices, also make optimization impossible.

Investors usually use ``backtesting'' to optimize their strategies\cite{harvey2015backtesting}. Backtesting is where an investors profits are estimated if they were trading at past market prices. As the past time evolution of market prices was fixed and unchanged, backtesting cannot handle the market impact. In fact, there is no method to exactly estimate the market impact before trading. Therefore, investors cannot estimate their earnings, including the effect of their market impacts. Furthermore, the changed market prices cause other investors to change their behaviors, which leads to more changes in the time evolution of market prices.

Investors implement optimized investment strategies to earn the best profits by backtesting. However, since market impacts change the market prices and other investors' behaviors from those in backtesting, such strategies can no longer earn the best profits. Even if investors re-optimize their strategies, they will again impact market prices, dooming this process of re-optimizing strategies to repeat forever. Therefore, the optimized parameter of an investment strategy is not stable. We refer to this phenomena as ``an optimization that is not stabilized'' or ``optimization instability.'' 

\begin{figure*}[t] 
\begin{center}
\includegraphics[scale=0.35]{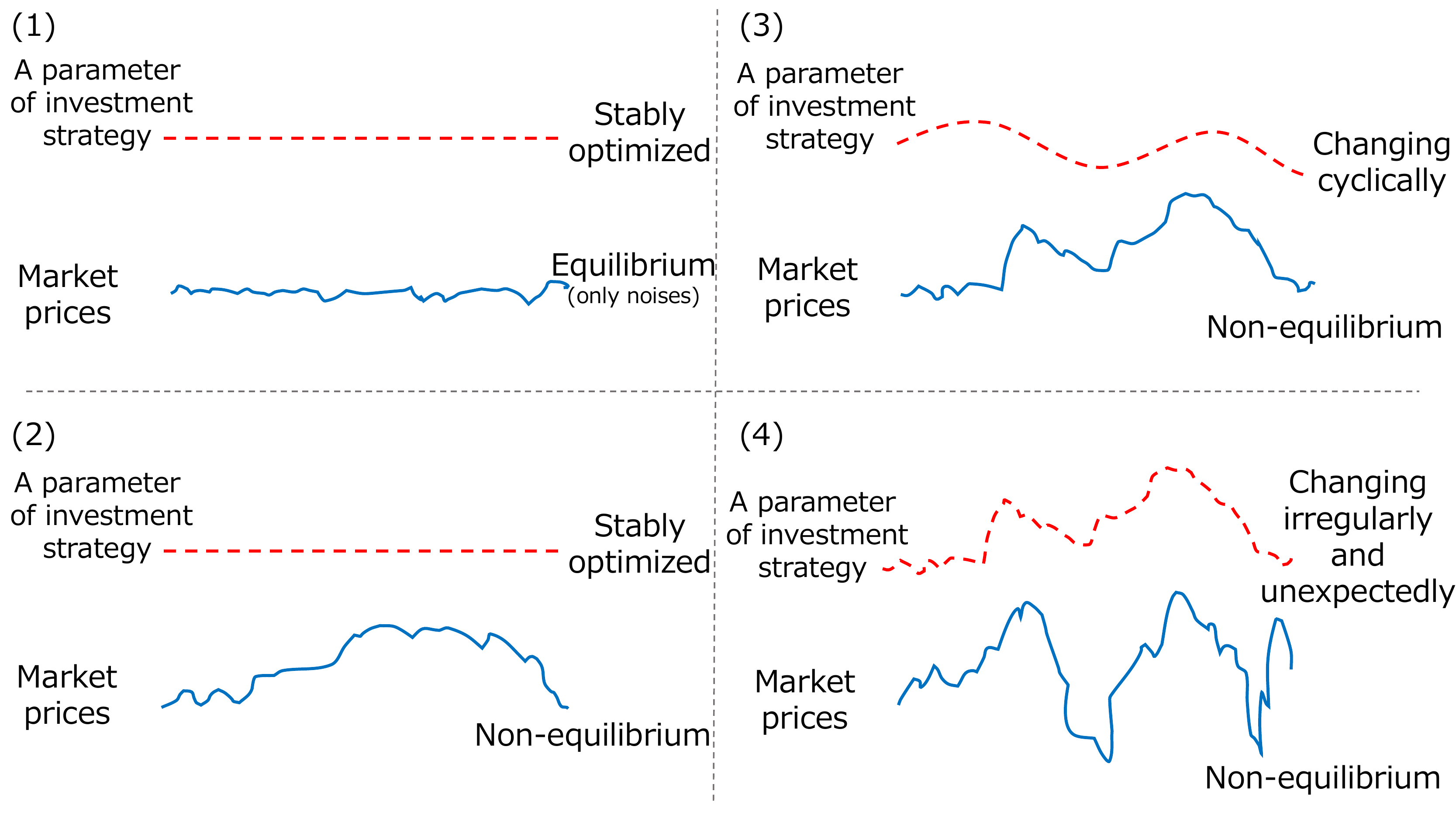}
\end{center}
\caption{Illustrations for (1): The parameter of an investment strategy is stably optimized and market prices reach equilibrium. (2): The parameter of an investment strategy is stably optimized but market prices are non-equilibrium. (3): The parameter of an investment strategy is unstably optimized and continues to change cyclically, and market prices are non-equilibrium. (4): The parameter of an investment strategy is unstably optimized and continues to change irregularly and unexpectedly, and market prices are non-equilibrium.}
\label{p07}
\end{figure*}

As Fig. \ref{p07} shows, the optimization instability is one level higher than ``non-equilibrium of market prices.'' In economics, market prices can either reach equilibrium or not, as shown in (1) and (2) in Fig. \ref{p07}, respectively. Many previous agent-based models for financial markets (artificial market models)\footnote{Excellent reviews include\cite{lebaron2006agent, chen2009agent, 7850016, mizuta2016SSRNrev, mizuta2019arxiv, mizuta2022aruka}. In the Appendix, we mention ``Basic concept for constructing a model'' and ``Validation of the model'' for an artificial market model on the basis of\cite{mizuta2019arxiv, mizuta2022aruka}.} showed that market prices do not reach equilibrium and continued to change not explained only by noise when there are many heterogeneous agents, as shown in (2) in Fig. \ref{p07}\cite{TAKAYASU1992127,lux1999scaling}.

In social sciences, the heterogeneousness of agents has become more important\cite{kirman2006heterogeneity}. In economics, heterogeneousness is also important especially after the financial crisis in 2009, and Stiglitz argued that the dynamic stochastic general equilibrium (DSGE) model that includes no heterogeneousness cannot explain real economics\cite{stiglitz2018modern}. In addition, there is also the argument that the DSGE model actually includes heterogeneousness\cite{christiano2018dsge}. In any case, heterogeneousness is one of the most important problem in economics.

Optimization instability causes not only market prices but also the parameters of investment strategies to continually change like those in (3) and (4) in Fig. \ref{p07}. A number of previous artificial market models implemented the learning process of agents in short-term price variations to modulate the parameters of strategies, and they contributed to the investigation of the mechanism of the market bubble and crash\cite{izumi1996artificial, arthur1997economy}. However, no artificial market model optimized investment strategies in which an agent searches one optimized parameter in a whole simulation run in economics and financial studies, and no model determined whether only market impacts lead to optimization instability.

Therefore, this study investigated, using an artificial market model, whether investors' inability to accurately estimate market impacts in their optimizations leads to optimization instability. Artificial market models include agents modeling investors' behaviors and show macro phenomena including market prices as a result of their interactions. Agent behaviors are simple but interact each other to cause complex macro phenomena, which are not a simple sum of the agent behaviors. Thus, an artificial market model can provide researchers with new knowledge. These micro-macro interactions sometimes give rise to strong phenomena called ``micro-macro feedback loops'' or ``positive feedback loops,'' in which micro processes strengthen macro phenomena, which in turn strengthen the micro processes, and these strengthenings continue as a loop. Effectively handling these loops is an advantage of an artificial market model. Loops caused by market impacts can potentially exist, which will lead to a more complex nature.

In this study, we built an artificial market model by adding two technical analysis strategy agents (TAs), which search one optimized parameter in a whole simulation run, to the prior model of Mizuta et al.\cite{mizuta2016ISAFM}. The TAs are a momentum TA (TA-m) and reversal TA (TA-r).

A technical analysis strategy uses a historical market return of prices. A momentum technical analysis strategy expects a positive (negative) return when the historical return is positive (negative). Conversely, a reversal technical analysis strategy expects a positive (negative) return when the historical return is negative (positive).

Chen et. al.\cite{chen2009agent} summarized previous studies on artificial market models and showed that models have to include technical strategies to replicate important stylized facts\footnote{A stylized fact is a term used in economics to refer to empirical findings that are so consistent (for example, across a wide range of instruments, markets, and time periods) that they are accepted as truth\cite{Sewell2006}.}. As comprehensively reviewed by Menkhoff and Taylor\cite{menkhoff2007}, many empirical questionnaire studies have found technical strategies in the real financial markets. An empirical data study gave the same conclusions\cite{yamamotopredictor}. Laboratory markets in experimental economics have greatly contributed to the aforementioned discussion. Parameter fitting of the artificial market model including technical strategies leads to similar results as those of the laboratory market\cite{Haruvy2006}.

After building the artificial market model, we investigated whether investors' inability to accurately estimate market impacts in their optimization leads to optimization instability.

\begin{figure}[t] 
\begin{center}
\includegraphics[scale=0.35]{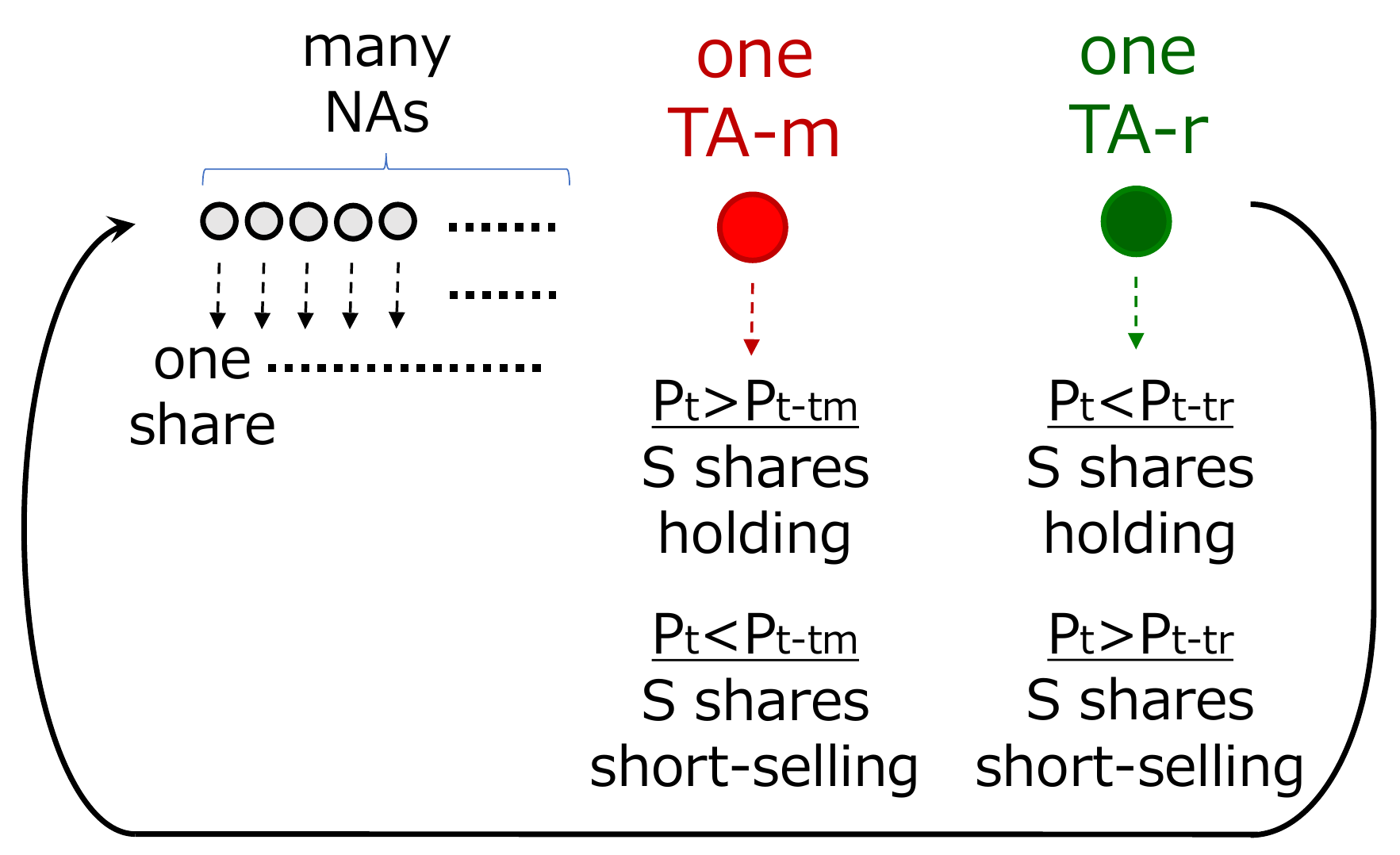}
\end{center}
\caption{Order of NA and TA trading}
\label{p03}
\end{figure}

\begin{figure}[t] 
\begin{center}
\includegraphics[scale=0.30]{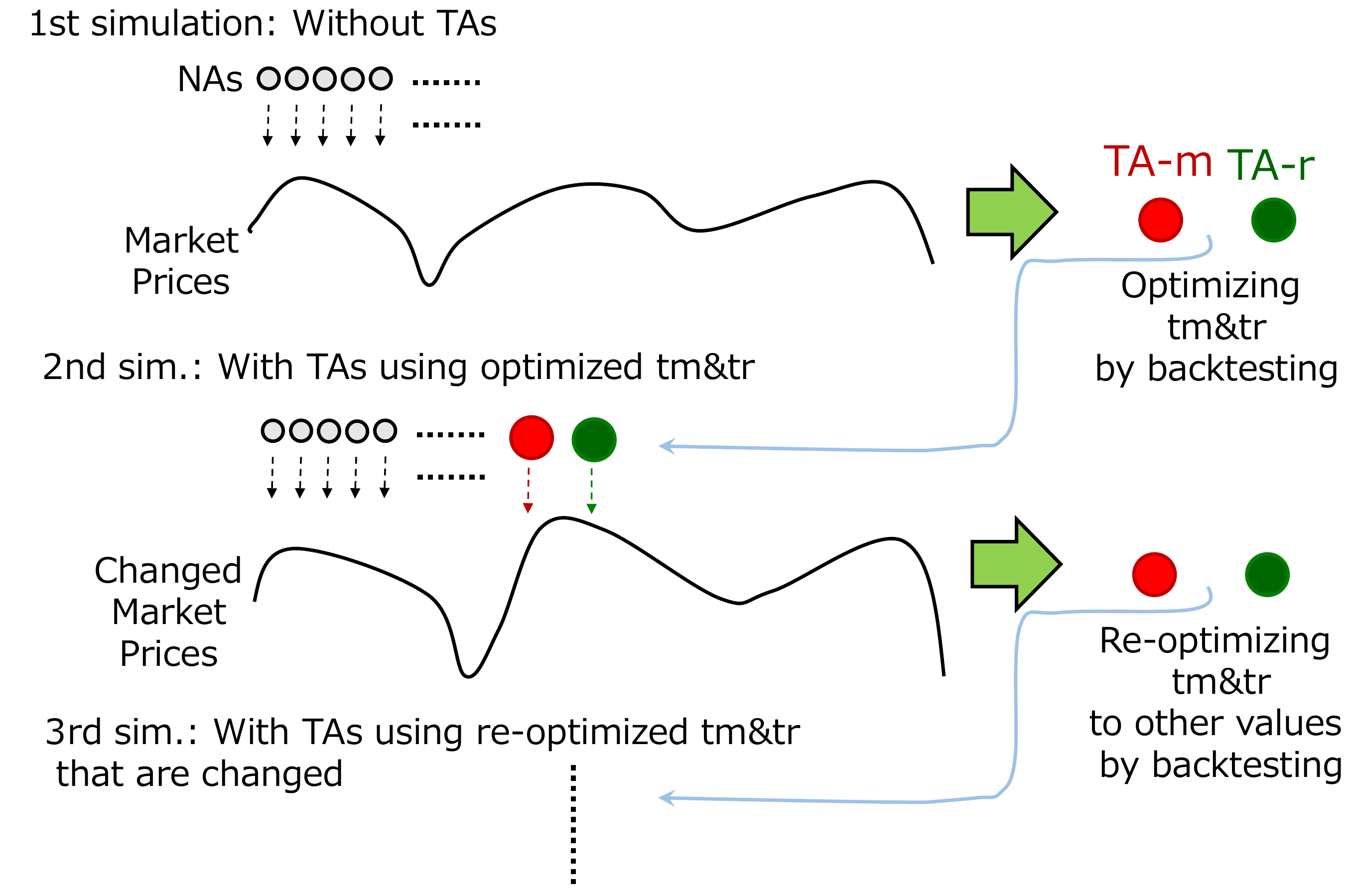}
\end{center}
\caption{TA learning}
\label{p04}
\end{figure}

\section{Model}
\label{s3}
The model of Chiarella and Iori\cite{chiarella2002simulation} is very simple but replicates long-term statistical characteristics observed in actual financial markets: a fat-tail and volatility clustering. In contrast, the model of Mizuta et al.\cite{mizuta2016ISAFM} replicates high-frequency micro structures, such as execution rates, cancel rates, and one-tick volatility, that cannot be replicated with that of Chiarella and Iori\cite{chiarella2002simulation}. Only fundamental and technical analysis strategies that exist generally for any market at any time\footnote{Many empirical studies using questionnaires found these strategies to be the majority, which are comprehensively reviewed by Menkhoff and Taylor\cite{menkhoff2007}. The empirical study using market data by Yamamoto\cite{yamamotopredictor} showed that investors are switching fundamental and technical analysis strategies.} are implemented into the agent model.

The simplicity of the model is very important for this study because unnecessary replication of macro phenomena leads to models that are overfitted and too complex. Such models prevent the understanding and discovery of mechanisms affecting price formation because of the increase in related factors. We explain the basic concept for constructing our artificial market model in the review article\cite{mizuta2019arxiv, mizuta2022aruka} and the Appendix ``Basic Concept for Constructing Model.''

In this study, the artificial market model was built by adding two technical agents (TAs), a momentum TA (TA-m) and reversal one (TA-r), to the prior model of Mizuta et al.\cite{mizuta2016ISAFM}.

This model contains one stock. The stock exchange uses a continuous double auction to determine the market price\cite{tse2015e}. In the auction mechanism, multiple buyers and sellers compete to buy and sell financial assets in the market, and transactions can occur at any time whenever an offer to buy and one to sell match. The minimum unit of a price change is $\delta P$. The buy-order and sell-order prices are rounder down and up to the nearest fraction, respectively.

The model includes $n$ normal agents (NAs), a TA-m, and TA-r. The NAs always place an order for only one share, while the TAs can place a number of orders at once. We implemented the variable ``tick time'' $t$, which increases by one when an NA orders.

\subsection{Normal agent (NA)}
To replicate the nature of price formation in actual financial markets, we introduced the NA to model a general investor. The number of NAs is $n$. The NAs can short sell freely. The holding positions are not limited, so the NAs can take an infinite number of shares for both long and short positions. Similarly to the process shown in Fig.\ref{p03}, at time $t=1$, NA No. $1$ first places an order to buy or sell its risk asset, and at $t=2,3,,,n$, NAs No. $2,3,,,n$ then respectively place buy or sell orders. At $t=n+1$, the TA-m and TA-r place orders when a trade condition is satisfied (see Section \ref{ss1}). After that, the model returns to the first NA and repeats this cycle. An NA determines the order price and buys or sells using a combination of fundamental and technical analysis strategies to form an expectation of a risk asset's return.

The expected return of agent $j$ for each risk asset at $t$ is
\begin{equation}
r^{t}_{e,j} = (w_{1,j} \ln{\frac{P_f}{P^{t-1}}} + w_{2,j}\ln{\frac{P^{t-1}}{P^{t-\tau _ j-1}}}+w_{3,j} \epsilon ^t _j )/\Sigma_i^3 w_{i,j} \label{eq1}
\end{equation}
where $w_{i,j}$ is the weight of term $i$ for agent $j$ and is independently determined by random variables uniformly distributed on the interval $(0,w_{i,max})$ at the start of the simulation for each agent. $\ln$ is the natural logarithm. $P_f$ is a fundamental value and is a constant. $P^t$ is a mid-price (the average of the highest buy-order price and the lowest sell-order price) at $t$, and $\epsilon ^t _ j$ is determined by random variables from a normal distribution with average $0$ and variance $\sigma _ \epsilon$ at $t$. $\tau_j$ is independently determined by random variables uniformly distributed on the interval $(1,\tau _{max})$ at the start of the simulation for each agent\footnote{When $t< \tau _ j$, the second term of Eq. (\ref{eq1}) is zero.}.

The first term in Eq. (\ref{eq1}) represents a fundamental strategy: the NA expects a positive return when the market price is lower than the fundamental value, and vice versa. The second term represents a technical analysis strategy using a historical return: the NA expects a positive return when the historical market return is positive, and vice versa. The third term represents noise.

After the expected return has been determined, the expected price is
\begin{equation}
P^t_{e,j}= P^{t-1} \exp{(r^t_{e,j})}.
\end{equation}

Order prices are scattered around the expected price $P^t_{e,j}$ to replicate many waiting limit orders. An order price $P^t_{o,j}$ is 
\begin{equation}
P^t_{o,j}=P^t_{e,j}+P_d(2\rho ^t _j-1),
\end{equation}
where $\rho ^t_j$ is determined by random variables uniformly distributed on the interval $(0,1)$ at $t$ and $P_d$ is a constant. This means that $P^t_{o,j}$ is determined by random variables uniformly distributed on the interval $(P^t_{e,j}-P_d, P^t_{e,j}+P_d)$ 

Whether the agent buys or sells is determined by the magnitude relationship between $P^t_{e,j}$ and $P^t_{o,j}$. When $P^t_{e,j}>P^t_{o,j}$, the NA places an order to buy one share, and when $P^t_{e,j}<P^t_{o,j}$, the NA places an order to sell one share\footnote{When $t<t_c$, to generate enough waiting orders, the agent places an order to buy one share when $P_f>P^t_{o,j}$, or to sell one share when $P_f<P^t_{o,j}$. \label{ft01}}. The remaining order is canceled $t_c$ after the order time.

\subsection{Technical agent (TA)}
\label{ss1}
Similarly to that shown in Fig. \ref{p03}, two TAs exist, a momentum TA (TA-m) and reversal TA (TA-r). After the NA No. $n$ places an order, the TA-m and TA-r place orders in this order when the following conditions are satisfied.\footnote{However, they place no order while $t<t_{max}$, which is the maximum of $tm$ and $tr$.}.

Trades of the TA-m are as follows. The TA-m places market buy-orders\footnote{When an agent orders to buy (sell), if there is a lower sell-order price (a higher buy-order price) than the agent's order, dealing immediately occurs. Such an order is called a ``market order.''} to hold $S$ shares when $P^t>P^{t-tm}$, and it places market sell-orders to short-sell $S$ shares (hold $-S$ shares) when $P^t<P^{t-tm}$, where $tm$ is a parameter of the TA-m trading strategy. The TA-m determines market tendency using market prices over the past $tm$.

Trades of the TA-r are as follows. The TA-r places market buy-orders to hold $S$ shares when $P^t<P^{t-tr}$, and it places market sell-orders to short-sell $S$ shares (hold $-S$ shares) when $P^t>P^{t-tr}$, where $tr$ is a parameter of the TA-r trading strategy. The TA-r determines market tendency using market prices over the past $tm$.

Note that if they already have the target shares, they place no orders.

\subsection{Learning of the TAs} 
\label{sss0}
Simulations are repeated like the processes shown in Fig. \ref{p04}. In the first simulation run, the case without the TAs is simulated. After that, the TAs optimize $tm$ and $tr$ to earn the best by backtesting and fixing the market prices of the simulation result using particle swarm optimization (PSO)\cite{488968} as mentioned in Section \ref{sss1}.

In the second simulation, the case with the TAs is simulated using $tm$ and $tr$ and exactly the same random numbers, $w_{i,j}$, $\tau _j$, $\epsilon ^t _j$, and $\rho ^t _j$ as those in the first simulation. Since the market prices are different from the first simulation, optimizing $tm$ and $tr$ to earn the best causes different results. Therefore, the TAs should re-optimize $tm$ and $tr$ using the new market prices generated by the second simulation.

The third simulation is simulated using the re-optimized parameters, in which the TAs re-optimize $tm$ and $tr$ again using the new market prices generated. In this way, the simulations are repeated to determine whether $tm$ and $tr$ converge to specific figures. Note that because random numbers $w_{i,j}$, $\tau _j$, $\epsilon ^t _j$, and $\rho ^t _j$ are fixed in all runs, when $tm$ and $tr$ are not changed, no parameters are changed, and the trades of all agents are also not changed. Thus, exactly the same simulation result is produced.

\subsection{Optimization of a strategy parameter}
\label{sss1}
After each simulation run, the TA-m optimizes $tm$ and the TA-r optimizes $tr$ independently between $t_{min}$ and $t_{max}$ to earn the best using PSO\cite{488968} in the following steps.

The TA-m and TA-r are independently investigated but in the same way. Here, the TA-m or TA-r is defined as the TA-x. In the PSO, there are many particles, and each particle has an input parameter and outputs result. In this study, a particle $k$ has an investment strategy parameter $t_k$ that corresponds to $tm$ or $tr$ as input and outputs a profit of backtesting when the TA-x uses $t_k$. The number of particles is $n_P$. $t_k$ is initially set $t_k=t_{min}+(t_{max}-t_{mix})(k/n_P)$. This means that $t_k$ is arranged at the same intervals from $t_{min}$ to $t_{max}$.

First, the profit of the TA-x with $t_k$ is calculated using backtesting. Then, $t_{kbest}$, which is the $t_k$ when the particle $k$ earns the best ever in the backtesting, is calculated. $t_{best}$, which is the $t_k$ when all particles earns the best ever in the backtesting, is also calculated. $t_k$ evolves as the following, 
\begin{equation}
\begin{array}{l}
\delta t_k \leftarrow w \delta t_k + c_1r_1(t_{best}-t_k)+c_2r_2(t_{kbest}-t_k) \\
t_k \leftarrow t_k+\delta t_k
\end{array}
\end{equation}
where $r_1$ and $r_2$ are determined by random variables uniformly distributed from $0$ to $1$ and $w$, $c_1$, and $c_2$ are constants.

The evolution is repeated $l_P$ times. $t_{best}$, which is the final output, is optimized by $tm$ or $tr$.

\begin{figure}[t] 
\begin{center}
\includegraphics[scale=0.35]{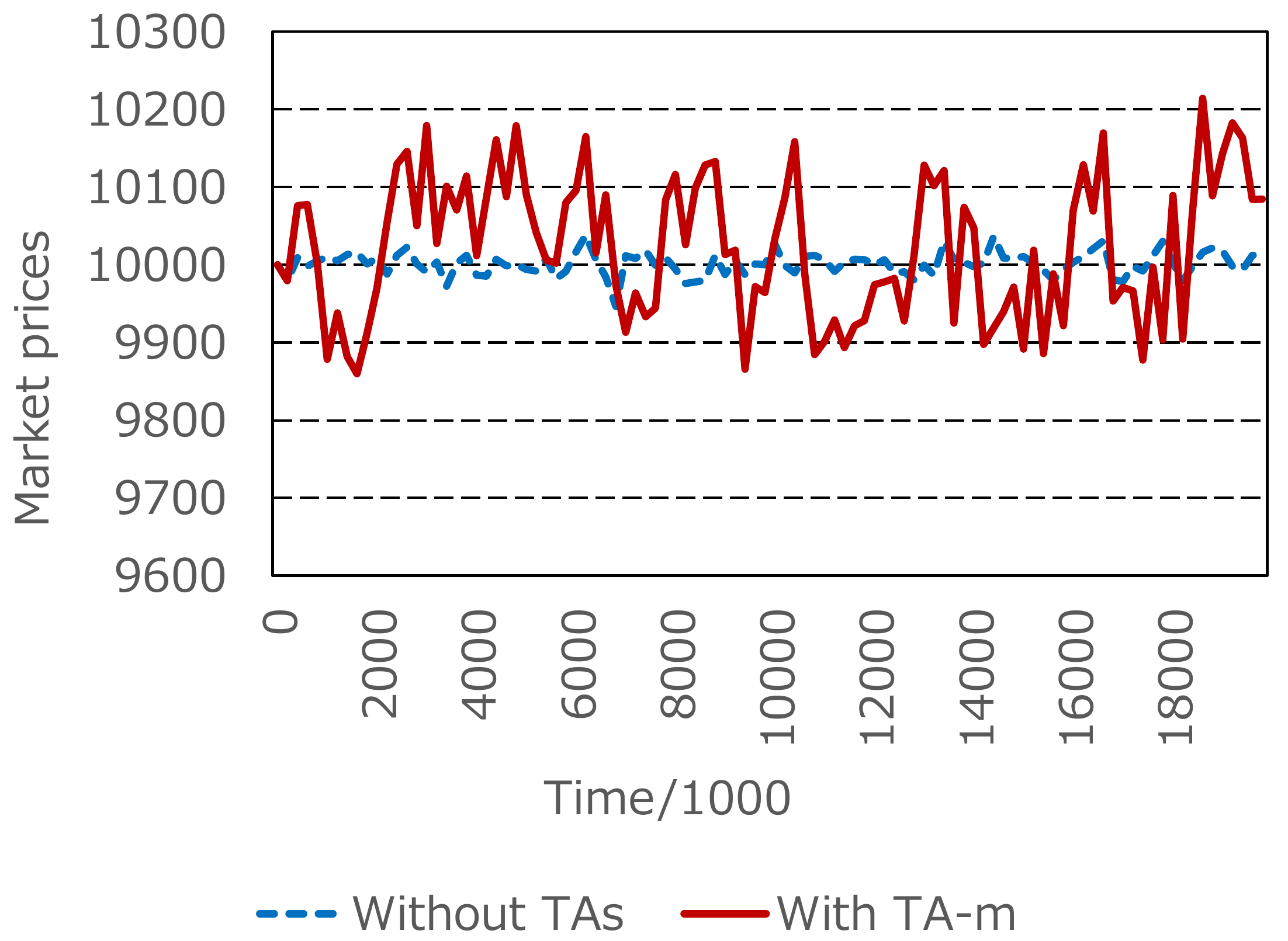}
\end{center}
\caption{Time evolution of prices when the TA-m using the $tm$ obtained after $50$ times optimizations exists and when it does not exist}
\label{e01}
\end{figure}

\begin{figure}[t] 
\begin{center}
\includegraphics[scale=0.35]{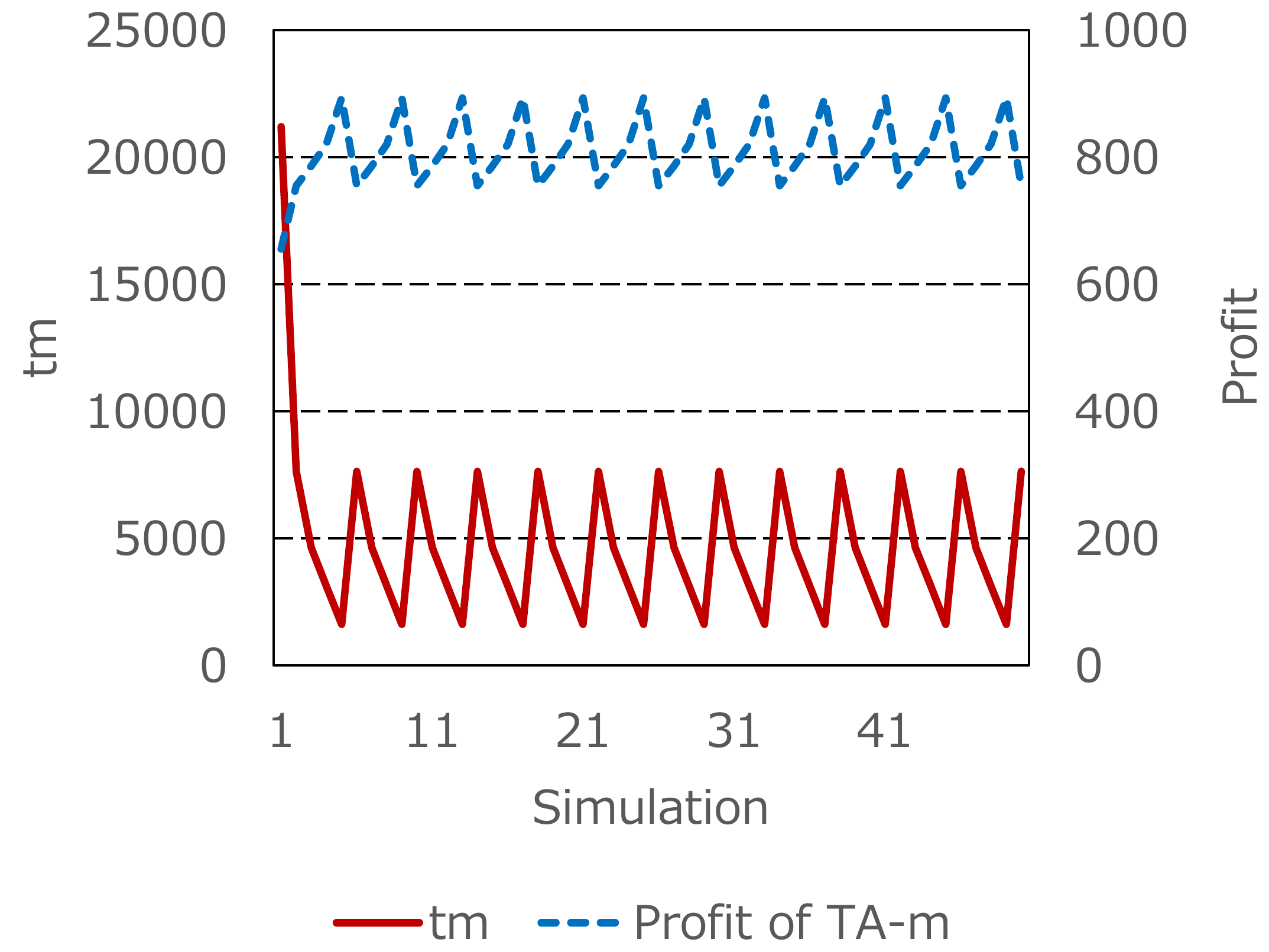}
\end{center}
\caption{$tm$ and profits of the TA-m for the simulations when the TA-m exists}
\label{e02}
\end{figure}

\begin{figure}[t] 
\begin{center}
\includegraphics[scale=0.35]{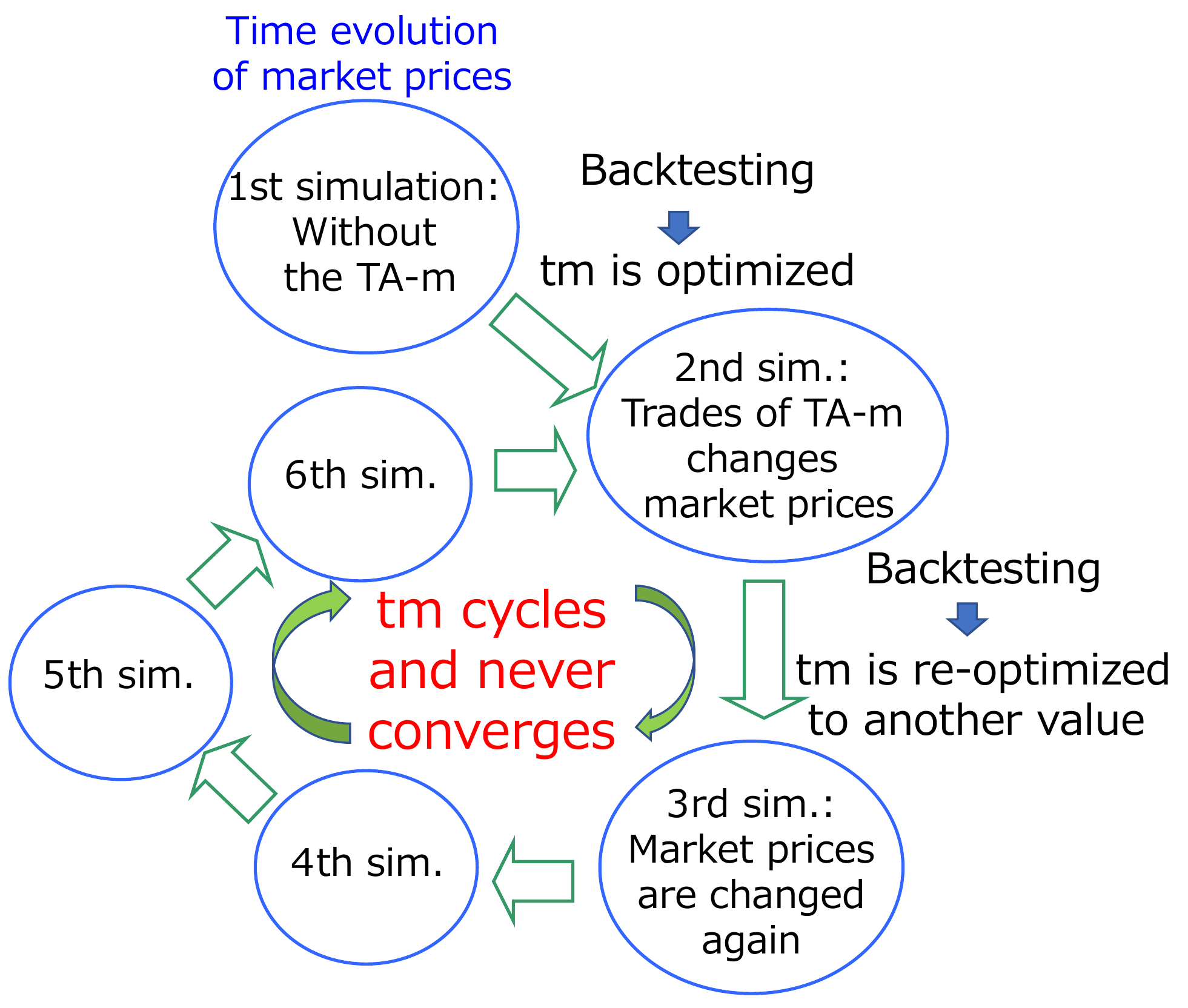}
\end{center}
\caption{Mechanism of $tm$ behavior}
\label{p01}
\end{figure}

\begin{figure}[t] 
\begin{center}
\includegraphics[scale=0.30]{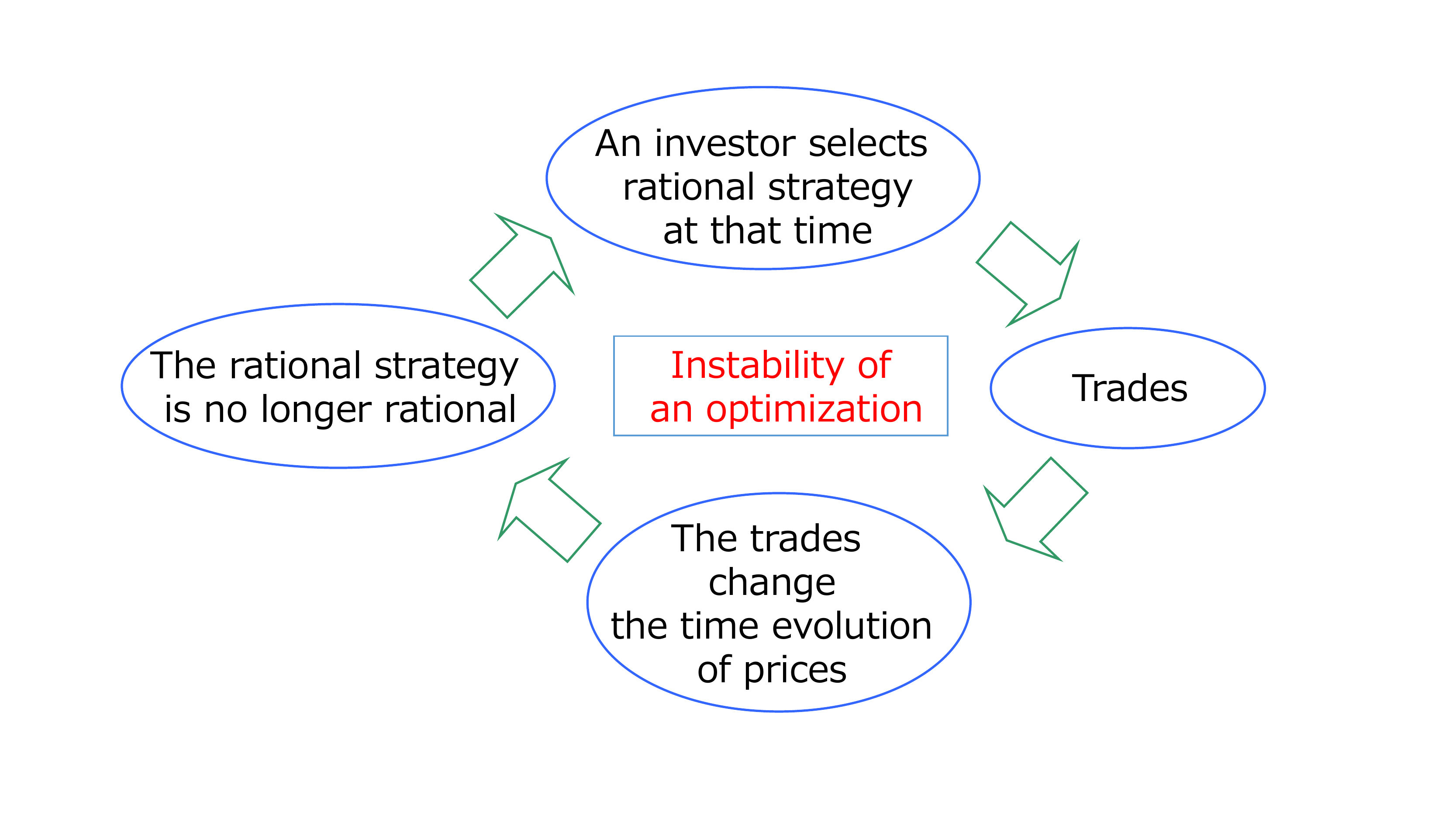}
\end{center}
\caption{Mechanism of instability by optimizing investment strategies}
\label{p05}
\end{figure}

\begin{figure}[t] 
\begin{center}
\includegraphics[scale=0.35]{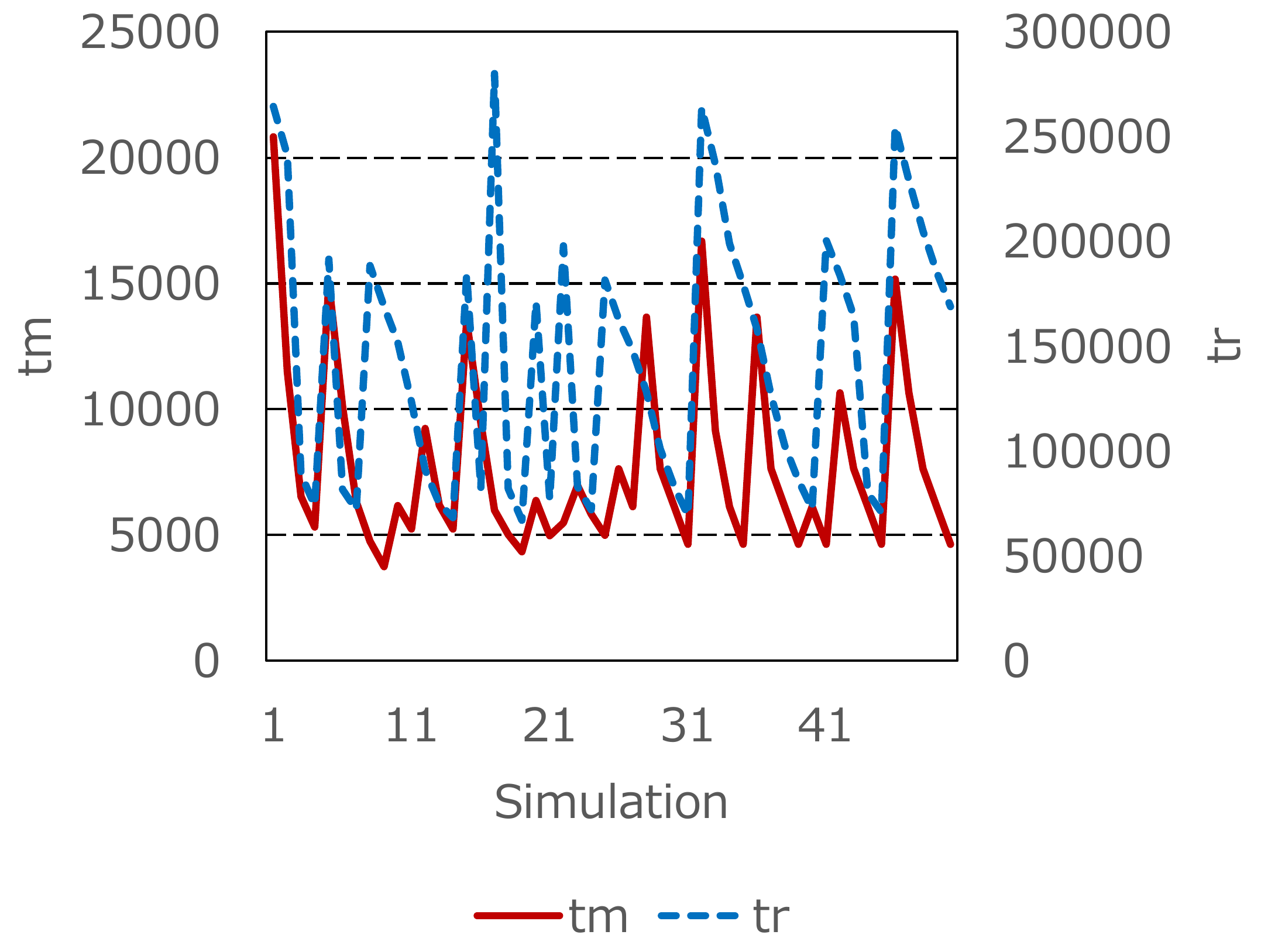}
\end{center}
\caption{$tm, tr$ for the simulations when both the TA-m and TA-r exist}
\label{e03}
\end{figure}

\begin{figure}[t] 
\begin{center}
\includegraphics[scale=0.35]{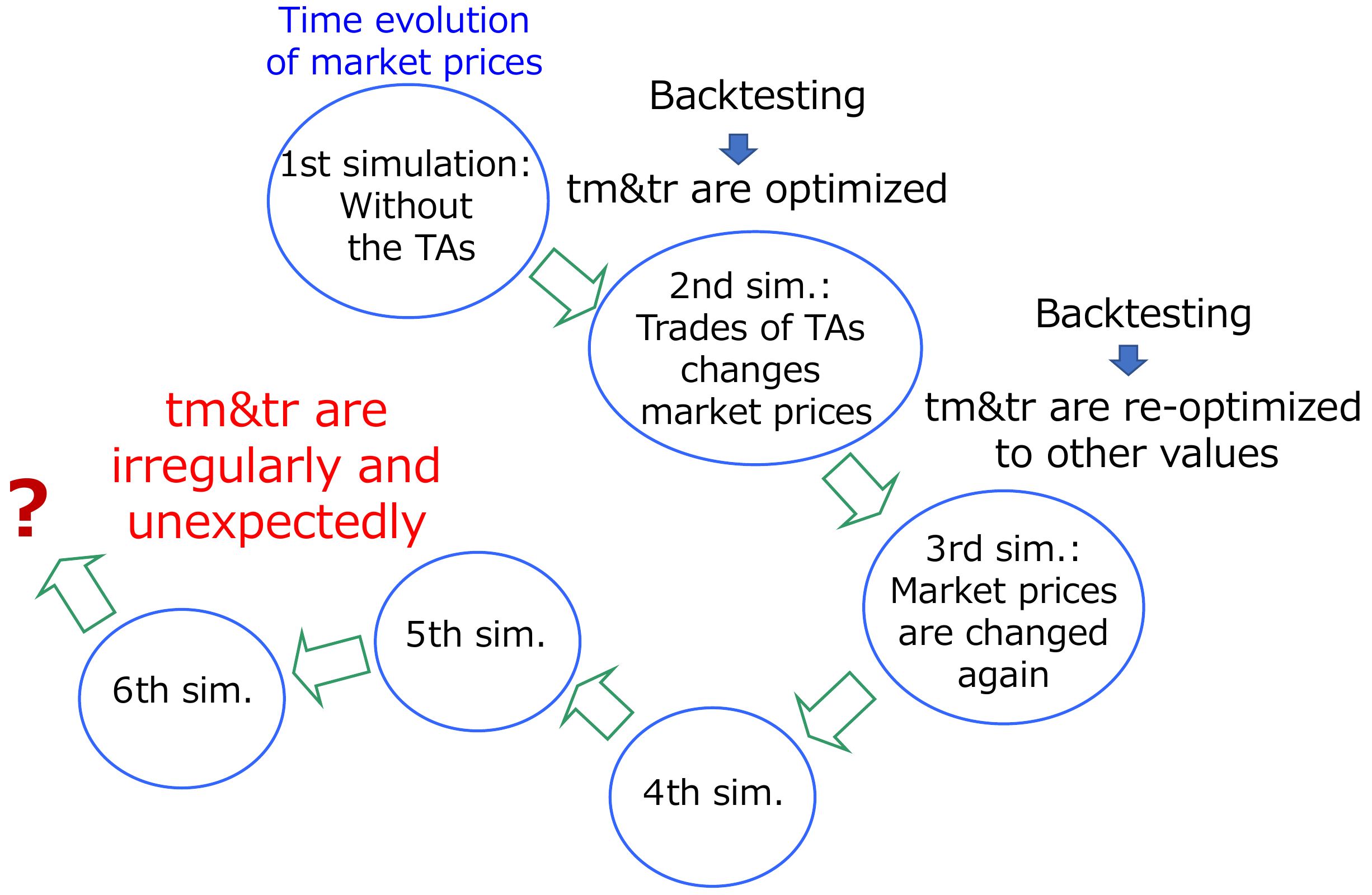}
\end{center}
\caption{Mechanism of $tm$ and $tr$ behavior}
\label{p02}
\end{figure}

\begin{figure}[t] 
\begin{center}
\includegraphics[scale=0.35]{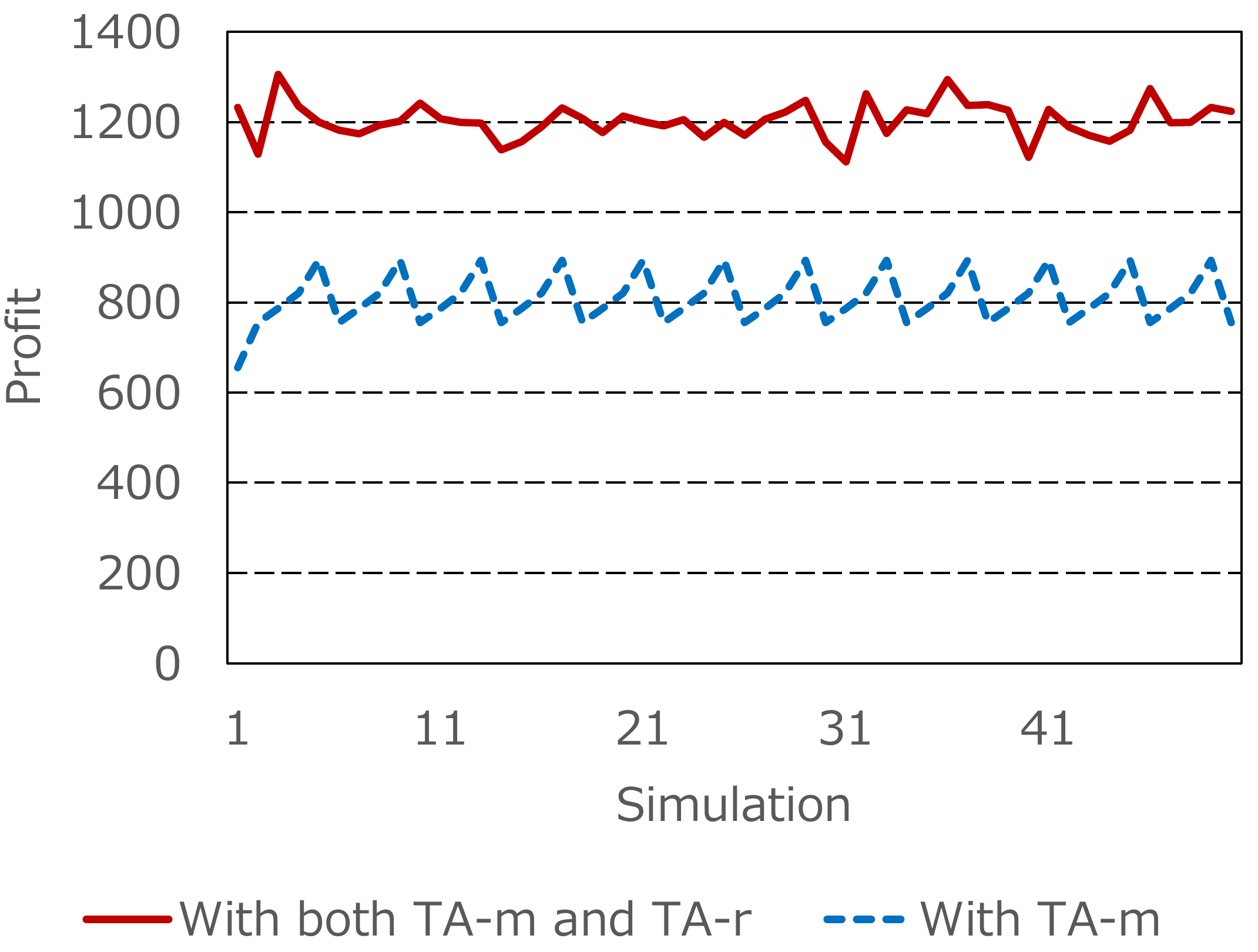}
\end{center}
\caption{Profits of TAs for the simulations where only the TA-m exists and both the TA-m and TA-r exist}
\label{e04}
\end{figure}

\section{Simulation results}
\label{s4}
In this study, we set\footnote{We explain how the model was validated in the Appendix ``Validation of the Model.''} $\delta P=0.01, P_f=10000$, and for the NAs, $n=1000, w_{1,max}=1, w_{2,max}=100, w_{3,max}=1, \tau _ {max}=10000, \sigma _ \epsilon = 0.03, P_d= 1000, t_c=10000$, and $P_{f}=10000$. For the TAs, we set $S=100$. The simulations ran to $t=t_e=20000000$. For the PSO, we set $n_P=200, l_P=50, t_{min}=100, t_{max}=300000, w=0.99, c_1=0.3, and c_2=0.3$.

Fig. \ref{e01} shows the time evolution of the prices when the TA-m using $tm$ obtained after the $50$th simulation exists and when it does not exist. The existence of the TA-m enlarged the price variation.

Fig. \ref{e02} shows $tm$ and the profits of the TA-m for the simulations when the TA-m exists. The profits were calculated as the final cash they have per $P_f$\footnote{Trades of the TAs leads to increased and decreased amounts of cash, while they initially have no cash and are permitted to have minus cash. Shares that the TAs have at the end of the simulation are calculated as $P_f$ per one share.}. The $tm$ never converged to a specific value, and this behavior corresponds to (3) in Fig. \ref{p07}.

Fig. \ref{p01} shows the mechanism of $tm$ behavior. First, the case without the TAs is simulated and the TA-m optimizes $tm$ to earn the best by backtesting and fixing the market prices of the simulation result. In the next simulation, because the TA-m trades using $tm$, the time evolution of market prices is changed, and the optimized $tm$ is also changed. After a number of repetitions of the optimizations, $tm$ returns to a certain historical value. After that, $tm$ changes in a cyclic manner. Strategies of traders never reach the equilibrium, and the optimized parameters of strategies are never fixed in time.

As mentioned in Section \ref{sss0}, because random numbers $w_{i,j}$, $\tau _j$, $\epsilon ^t _j$, and $\rho ^t _j$ are fixed in all runs, when $tm$ is not changed, no parameters are changed, and the trades of all agents are exactly the same. Thus, exactly the same simulation result is produced and the time evolution of prices is the same. Therefore, the optimized $tm$ never stabilizes even when the whole situation other than the TA-m is exactly same.

This means that even if all other traders are fixed, only one investor optimizing his/her strategy using backtesting leads to the time evolution of market prices becoming unstable. Financial markets are essentially unstable, and naturally, investment strategies are not able to be fixed. The reason is that, as shown in Fig. \ref{p05}, even when one investor selects a rational strategy at that time, it changes the time evolution of prices, it becomes no longer rational, another strategy becomes rational, and the process repeats.

Fig. \ref{e03} shows $tm, tr$ for the simulations when both the TA-m and TA-r exist. The changes of $tm, tr$ were irregular and unexpected, and the cyclic feature was only present in the case where only the TA-m exists, as shown in Fig. \ref{e02}. This feature corresponds to (4) in Fig. \ref{p07}. The time evolution of market prices produced by investment strategies having such unstable parameters is highly unlikely to be predicted and have stable laws written by equations. This nature makes us suspect that financial markets include the principle of natural uniformity and indicates the difficulty of building an equation model explaining the time evolution of prices.

Fig. \ref{p02} shows the mechanism of $tm, tr$ behavior. Similarly to the case where only the TA-m exists (Fig. \ref{p01}), when the TAs optimize $tm, tr$, the time evolution of market prices and the optimal $tm, tr$ are changed. However, both $tm and tr$ do not return to certain historical values compared with the case where only the TA-m exists, so the changes of $tm, tr$ are irregular and unexpected.

Fig. \ref{e04} shows the profits of the TA-m for the simulations where only the TA-m exists and both the TA-m and TA-r exist. When both the TA-m and TA-r existed, the TA-m earned more than that without the TA-r. A profit of one strategy is not stolen by the trades of another strategy, but both strategies increase their profits. The result is consistent with the previous study showing when an opposite strategy exists, another strategy can earn more without the opposite strategy using an artificial market model\cite{mizuta2021BESC}\footnote{In\cite{mizuta2021BESC}, the parameters of each strategy are fixed and the learning process is not implemented.}.

\section{Summary}
\label{s5}
In this study, we built an artificial market model by adding technical analysis strategy agents (TAs), which search one optimized parameter in a whole simulation run, to the prior model of Mizuta et al.\cite{mizuta2016ISAFM}. The TAs are a momentum TA (TA-m) and reversal TA (TA-r), and we investigated whether investors' inability to accurately estimate market impacts in their optimizations leads to optimization instability.

When only the TA-m exists, the parameter of investment strategy never converged to a specific value, but changed in a cyclic manner. This means that even if all other traders are fixed, only one investor optimizing his/her strategy using backtesting leads to the time evolution of market prices becoming unstable. Financial markets are essentially unstable, and naturally, investment strategies are not able to be fixed. The reason is that even when one investor selects a rational strategy at that time, it changes the time evolution of prices, it becomes no longer rational, another strategy becomes rational, and the process repeats.

When both the TA-m and TA-r exist, the parameters of investment strategies were changing irregularly and unexpectedly, and the cyclic feature was only present in the case where only the TA-m exists. Optimization instability is one level higher than ``non-equilibrium of market prices.'' Therefore, the time evolution of market prices produced by investment strategies having such unstable parameters is highly unlikely to be predicted and have stable laws written by equations. This nature makes us suspect that financial markets include the principle of natural uniformity and indicates the difficulty of building an equation model explaining the time evolution of prices.

\section{Appendix}
\subsection{Basic concept for constructing a model}
An artificial market model, which is a kind of agent-based model, can be used to investigate situations that have never occurred, handle regulation changes that have never been made, and isolate the pure contribution of these changes to price formation and liquidity\cite{mizuta2019arxiv, mizuta2022aruka}. These are the advantages of an artificial market simulation.

However, the outputs of this simulation would not be accurate or be credible forecasts of the actual future. The simulation needs to reveal possible mechanisms that affect price formation through many simulation runs, e.g., searching for parameters or purely comparing the before and after states of changes. The possible mechanisms revealed by these runs provide new intelligence and insight into the effects of the changes in price formations in actual financial markets. Other methods of study, e.g., empirical studies, would not reveal such possible mechanisms.

Artificial markets should replicate the macro phenomena that exist generally for any asset at any time. Price variation, which is a kind of macro phenomenon, is not explicitly modeled in artificial markets. Only micro processes, agents (general investors), and price determination mechanisms (financial exchanges) are explicitly modeled. Macro phenomena emerge as the outcome of interactions from micro processes. Therefore, the simulation outputs should replicate existing macro phenomena to generally prove that simulation models are probable in actual markets.

However, it is not the primary purpose for an artificial market to replicate specific macro phenomena only for a specific asset or period. Unnecessary replication of macro phenomena leads to models that are overfitted and too complex. Such models would prevent us from understanding and discovering mechanisms that affect price formation because the number of related factors would increase.

In addition, artificial market models that are too complex are often criticized because they are very difficult to evaluate\cite{chen2009agent}. A model that is too complex not only would prevent us from understanding mechanisms but also could output arbitrary results by overfitting too many parameters. It is more difficult for simpler models to obtain arbitrary results, and these models are easier to evaluate.

Therefore, we constructed an artificial market model that is as simple as possible and does not intentionally implement agents to cover all the investors who would exist in actual financial markets.

As Michael Weisberg mentioned\cite{Weisberg2012}, ``Modeling, (is) the indirect study of real-world systems via the construction and analysis of models.'' ``Modeling is not always aimed at purely veridical representation. Rather, they worked hard to identify the features of these systems that were most salient to their investigations.''

Therefore, effective models are different depending on the phenomena they focus on. Thus, our model is effective only for the purpose of this study and not for others. The aim of our study is to understand how important properties (behaviors, algorithms) affect macro phenomena and play a role in the financial system rather than representing actual financial markets precisely.

The aforementioned discussion holds not only for artificial markets but also for agent-based models used in fields other than financial markets. For example, Thomas Schelling, who received the Nobel Prize in economics, used an agent-based model to discuss the mechanism of racial segregation. The model was built very simply compared with an actual town to focus on the mechanism\cite{Schelling2006}. While it was not able to predict the segregation situation in the actual town, it was able to explain the mechanism of segregation as a phenomenon.

Harry Stevens, a newspaper writer, simulated an agent-based model to explain the spread of COVID-19 and to determine how to prevent infection\cite{Stevens2020}. The model was too simple to replicate the real world, but its simplicity enabled it to reveal the mechanism behind the spread.

Michael Weisberg studied what mathematical and simulation models are in the first place and cited the example of a map\cite{Weisberg2012}. Needless to say, a map models geographical features on the way to a destination. With a simple map, we can easily understand the way to the destination. However, while a satellite photo replicates actual geographical features very well, we cannot easily find the way to the destination.

The title page of Michael Weisberg's book\cite{Weisberg2012} cited a passage from a short story by Jorge Borges\cite{Borges}, ``In time, those Unconscionable Maps no longer satisfied, and the Cartographers Guilds struck a Map of the Empire whose size was that of the Empire, and which coincided point for point with it...In the Deserts of the West, still today, there are Tattered Ruins of that Map, inhabited by Animals and Beggars.'' The story in which a map was enlarged to the same size as the real Empire to become the most detailed of any map is an analogy to that too detailed a model is not useful. This story give us one of the most important lessons for when we build and use any model.

\begin{table}[t]
\caption{Stylized facts without the TAs}
\begin{center}
 \begin{tabular}{ccr}
 \multicolumn{2}{c}{kurtosis or returns} & $18.341$ \\ \hline
 & lag & \\
 & 1 & $0.060$ \\
 autocorrelation & 2 & $0.050$ \\
 coefficient for & 3 & $0.047$ \\
 square returns & 4 & $0.038$ \\
 & 5 & $0.036$ 
 \end{tabular}
\label{t0}
\end{center}
\end{table}

\subsection{Validation of the model}
In many previous artificial market studies, the models were validated to determine whether they could explain stylized facts, such as a fat-tail or volatility clustering\cite{lebaron2006agent,chen2009agent, mizuta2019arxiv, mizuta2022aruka}. A fat-tail means that the kurtosis of price returns is positive. Volatility clustering means that square returns have a positive autocorrelation, which slowly decays as its lag becomes longer.

Many empirical studies, e.g., that of Sewell\cite{Sewell2006}, have shown that both stylized facts (fat-tail and volatility clustering) exist statistically in almost all financial markets. Conversely, they also have shown that only the fat-tail and volatility clustering are stably observed for any asset and in any period because financial markets are generally unstable. This leads to the conclusion that an artificial market should replicate macro phenomena that exist generally for any asset at any time, fat-tails, and volatility clustering. This is an example of how empirical studies can help an artificial market model.

The kurtosis of price returns and the autocorrelation of square returns are stably and significantly positive, but the magnitudes of these values are unstable and very different depending on the asset and/or period. Very broad magnitudes of about $1 \sim 100$ and about $0 \sim 0.2$, respectively, have been observed\cite{Sewell2006}.

For the aforementioned reasons, an artificial market model should replicate these values as significantly positive and within a reasonable range. It is not essential for the model to replicate specific values of stylized facts because the values of these facts are unstable in actual financial markets.

Table \ref{t0} lists the statistics showing the stylized facts, kurtosis of price returns for $100$ tick times ($\ln(P^t/P^{t-100})$), and autocorrelation coefficient for square returns for $100$ tick times without TAs. This shows that this model replicated the statistical characteristics, fat-tails, and volatility clustering observed in real financial markets.

\bibliographystyle{IEEEtran}
\bibliography{ref}

\end{document}